\documentclass{aastex631}
\usepackage{natbib}
\usepackage{amsmath}
\usepackage{verbatim}
\usepackage{amssymb}
\usepackage{hyperref}
\usepackage{float}
\usepackage{graphicx}
\usepackage{mathtools}
\usepackage{empheq}
\usepackage{appendix}
\usepackage{microtype}
\usepackage{commath}
\usepackage{bm}
\usepackage{bold-extra}

\DeclareGraphicsExtensions{.pdf,.png,.jpg}

\newcommand{\bq}{\begin{equation}} 
\newcommand{\eq}{\end{equation}} 
 
\newcommand{\beq}{\begin{equation}}
\newcommand{\eeq}{\end{equation}}
\newcommand{\beqa}{\begin{eqnarray}}
\newcommand{\eeqa}{\end{eqnarray}}

\newcommand{\PL}{$P$--$L$\ }
\newcommand{\PLs}{$P$--$L$}

\defcitealias{Riess:2023}{R23}

\defcitealias{Riess:2022}{R22}

\received{Dec 21, 2023}
\accepted{ApJ Letters, Jan 3, 2024}

\shorttitle{JWST Cycle 1}
\shortauthors{Riess et al.}

\begin{document}

\title{JWST Observations Reject Unrecognized Crowding of Cepheid Photometry \\ as an Explanation for the Hubble Tension at 8$\sigma$ Confidence} 

\author[0000-0002-6124-1196]{Adam G.~Riess}
\affiliation{Space Telescope Science Institute, 3700 San Martin Drive, Baltimore, MD 21218, USA}
\affiliation{Department of Physics and Astronomy, Johns Hopkins University, Baltimore, MD 21218, USA}

\author[0000-0002-5259-2314]{Gagandeep S. Anand}
\affiliation{Space Telescope Science Institute, 3700 San Martin Drive, Baltimore, MD 21218, USA}

\author[0000-0001-9420-6525]{Wenlong Yuan}
\affiliation{Department of Physics and Astronomy, Johns Hopkins University, Baltimore, MD 21218, USA}

\author{Stefano Casertano}
\affiliation{Space Telescope Science Institute, 3700 San Martin Drive, Baltimore, MD 21218, USA}

\author{Andrew Dolphin}
\affiliation{Raytheon, 1151 E. Hermans Road, Tucson, AZ 85706, USA}

\author[0000-0002-1775-4859]{Lucas M.~Macri}
\affiliation{NSF's NOIRLab, 950 N Cherry Ave, Tucson, AZ 85719, USA}

\author[0000-0003-3889-7709]{Louise Breuval}
\affiliation{Department of Physics and Astronomy, Johns Hopkins University, Baltimore, MD 21218, USA}

\author[0000-0002-4934-5849]{Dan Scolnic}
\affiliation{Department of Physics, Duke University, Durham, NC 27708, USA}

\author{Marshall Perrin}
\affiliation{Space Telescope Science Institute, 3700 San Martin Drive, Baltimore, MD 21218, USA}

\author[0000-0001-8089-4419]{Richard I.~Anderson}
\affiliation{Institute of Physics, \'Ecole Polytechnique F\'ed\'erale de Lausanne (EPFL),\\ Observatoire de Sauverny, 1290 Versoix, Switzerland}

\begin{abstract}
We present high-definition observations with the {\it James Webb Space Telescope} of $>$1000 Cepheids in a geometric anchor of the distance ladder, NGC$\,$4258, and in 5 hosts of 8 Type Ia supernovae, a far greater sample than previous studies with {\it JWST}.  These galaxies individually contain the largest samples of Cepheids, an average of $>$150 each, producing the strongest statistical comparison to those previously measured with the {\it Hubble Space Telescope} in the NIR.  They also span the distance range of those used to determine the Hubble constant with {\it HST}, allowing us to search for a distance-dependent bias in {\it HST} measurements.  The superior resolution of {\it JWST} negates crowding noise, the largest source of variance in the NIR Cepheid Period-Luminosity relations (Leavitt laws) measured with {\it HST}.  Together with the use of two-epochs to constrain Cepheid phases and three filters to remove reddening, we reduce the dispersion in the Cepheid \PL relations by a factor of 2.5.  We find no significant difference in the mean distance measurements determined from {\it HST} and {\it JWST}, with a formal difference of -0.01$\pm$0.03 mag.  This result is independent of zeropoints and analysis variants including metallicity dependence, local crowding, choice of filters, and slope of the relations.  We can reject the hypothesis of unrecognized crowding of Cepheid photometry from {\it HST} that grows with distance as the cause of the ``Hubble Tension'' at 8.2$\sigma$, i.e., greater confidence than that of the Hubble Tension itself.   We conclude that errors in photometric measurements of Cepheids across the distance ladder do not significantly contribute to the Tension.

\end{abstract}

\section{Introduction} \label{sec:intro}

In the past decade, an intriguing and persistent discrepancy referred to as the ``Hubble Tension"\footnote{Tension may refer to the discrepancy between measures or to the feeling it produces as expressed by \cite{Verde:2023}: ``The research community has been actively looking for deviations from  $\Lambda$CDM for two decades; the one we might have found makes us wish we could put the genie back in the bottle."} has been apparent at high significance ($>$5$\sigma$) between the Hubble constant (H$_0$) directly measured from redshifts and distances, which are independent of cosmological models, and the same 
parameter derived from the $\Lambda$CDM model calibrated in the early Universe \citep[for a recent review, see][]{Verde:2023}.  

The most significant disparity arises from the strongest constraints.  These come from measurements of 42 local Type Ia supernovae (SNe~Ia) calibrated by Cepheid variables, yielding H$_0=73.0 \pm 1.0$ km/s/Mpc (SH0ES Collaboration, \citealt{Riess:2022}, hereafter \citetalias{Riess:2022}), compared to the analysis of Planck observations of the Cosmic Microwave Background \citep{Planck:2018}, predicting H$_0=67.4 \pm 0.5$ km/s/Mpc in conjunction with $\Lambda$CDM. Cepheids are the preferred primary distance indicators in these studies due to the Leavitt Law \citep[\PL\ relation;][]{Leavitt:1912}, their extraordinary luminosity (M$_H\sim-$7 mag at a period of 30 days), intrinsic precision (approximately 3\% in distance per star), reliable identification based on periodicity and light curve shape \citep{Hertzsprung:1926}, and comprehensive understanding \citep[since][]{Eddington:1917}. They also serve as the best-calibrated distance indicators accessible in the largest volume of SN~Ia hosts ($D \sim $ 50 Mpc), thanks to the consistent use of a single stable instrument, {\it HST} WFC3 UVIS+IR, by the SH0ES Team in measurements within SN hosts and in several independent geometric anchors: the megamaser host NGC~4258 \citep{Reid:2019}, the Milky Way \citep[through parallaxes, now including {\it Gaia} EDR3;][]{Gaia-Collaboration:2020}, and the Magellanic Clouds \citep[via detached eclipsing binaries;][]{Pietrzynski:2019}. Near-infrared (NIR) observations are crucial to mitigating the impact of dust, a challenge faced by many cosmic probes.  Yet, the modest NIR resolution of {\it HST}, $\sim$0$\farcs$1, has limited the inherent precision of individual Cepheid measurements due to the effects of crowding in nearby galaxies.   As noted by \citet{Freedman:2019}, ``[p]ossibly the most significant challenge for Cepheid measurements beyond 20 Mpc is crowding and blending from redder (RGB and AGB) disk stars, particularly for near-infrared H-band measurements [...]”.  

The {\it James Webb Space Telescope} ({\it JWST}) provides new capabilities to scrutinize and refine the strongest observational evidence contributing to the Tension. 
Specifically, the significantly greater resolution of {\it JWST} over {\it HST} has greatly reduced---in practical terms, almost eliminated---the main source of noise in near-infrared photometry of Cepheid variables observed in the hosts of nearby SNe~Ia. The resolution of {\it JWST} provides the ability to cleanly separate these vital standard candles from surrounding photometric ``chaff.''   This study extends the scope of {\it JWST} measurements of Cepheids along the distance ladder, building upon measurements in one SN~Ia host (NGC$\,$5584) and the distance scale anchor NGC$\,$4258 (\citealt{Riess:2023}, hereafter \citetalias{Riess:2023}).  Here we present a greatly expanded sample of such measurements that doubles its distance range to span the full range of distances of nearby SN calibrators (the second rung of the distance ladder) and triples the Cepheid sample size, while raising the number of SN hosts studied from 1 to 5 and the number of SN~Ia calibrated from 1 to 8. In \S 2 we present the observations, in \S 3 their analysis and in \S 4 we discuss their interpretation.

\begin{deluxetable*}{lrllllcrrr}[t]
\tabletypesize{\scriptsize}
\tablecaption{Observation Log\label{tb:obs}}
\tablewidth{0pt}
\tablehead{
\multicolumn{1}{c}{Date} & \multicolumn{1}{c}{MJD} & \multicolumn{1}{c}{Epoch} & \multicolumn{1}{c}{Exposure$^a$} & \multicolumn{1}{c}{Filter1} & \multicolumn{1}{c}{Filter2} & \multicolumn{1}{c}{Exp. time [s]} & \multicolumn{1}{c}{RA (J2000)} & \multicolumn{1}{c}{Dec (J2000)} & \multicolumn{1}{c}{Orientation}
}
\startdata
2023-06-30 &   60125.43 &    N1559e1 &            001001\_02101\_* &  F090W &  F277W &    418.7$\times4$ &     64.39918 &    -62.78429 &    216.0 \\
2023-06-30 &   60125.46 &    N1559e1 &            001001\_04101\_* &  F150W &  F277W &    526.1$\times4$ &     64.39918 &    -62.78429 &    216.0 \\
2023-07-15 &   60140.66 &    N1559e2 &            002001\_03101\_* &  F090W &  F277W &    418.7$\times4$ &     64.39906 &    -62.78429 &    221.0 \\
2023-07-15 &   60140.69 &    N1559e2 &            002001\_03103\_* &  F150W &  F277W &    526.1$\times4$ &     64.39906 &    -62.78429 &    221.0 \\
2023-07-07 &   60132.13 &    N5643e1 &            011001\_02101\_* &  F090W &  F277W &    311.4$\times4$ &    218.16850 &    -44.17334 &     91.7 \\
2023-07-07 &   60132.15 &    N5643e1 &            011001\_04101\_* &  F150W &  F277W &    418.7$\times4$ &    218.16850 &    -44.17334 &     91.7 \\
2023-07-22 &   60147.17 &    N5643e2 &            012001\_03101\_* &  F090W &  F277W &    311.4$\times4$ &    218.18105 &    -44.12627 &     96.7 \\
2023-07-22 &   60147.19 &    N5643e2 &            012001\_05101\_* &  F150W &  F277W &    418.7$\times4$ &    218.18105 &    -44.12627 &     96.7 \\
2023-08-02 &   60158.89 &    N1448e1 &            013001\_02101\_* &  F090W &  F277W &    418.7$\times4$ &     56.16443 &    -44.61610 &    251.0 \\
2023-08-02 &   60158.91 &    N1448e1 &            013001\_04101\_* &  F150W &  F277W &    526.1$\times4$ &     56.16443 &    -44.61610 &    251.0 \\
2023-08-18 &   60174.19 &    N1448e2 &            014001\_03101\_* &  F090W &  F277W &    418.7$\times4$ &     56.17600 &    -44.66334 &    256.0 \\
2023-08-18 &   60174.21 &    N1448e2 &            014001\_05101\_* &  F150W &  F277W &    526.1$\times4$ &     56.17600 &    -44.66334 &    256.0 \\
2023-06-28 &   60123.27 &    N5468e1 &            007001\_02101\_* &  F090W &  F277W &    204.0$\times6$ &    211.66669 &     -5.40969 &    114.6 \\
2023-06-28 &   60123.29 &    N5468e1 &            007001\_02103\_* &  F150W &  F277W &    472.4$\times6$ &    211.66669 &     -5.40969 &    114.6 \\
2023-07-14 &   60139.55 &    N5468e2 &            008001\_02101\_* &  F090W &  F277W &    311.4$\times5$ &    211.66745 &     -5.41007 &    115.6 \\
2023-07-14 &   60139.57 &    N5468e2 &            008001\_02103\_* &  F150W &  F277W &    526.1$\times5$ &    211.66745 &     -5.41007 &    115.6
\enddata
\tablecomments{$a$: All exposures start with jw01685.}
\end{deluxetable*}
\begin{figure}[t]
\begin{center}
\includegraphics[width=0.99\textwidth]{ 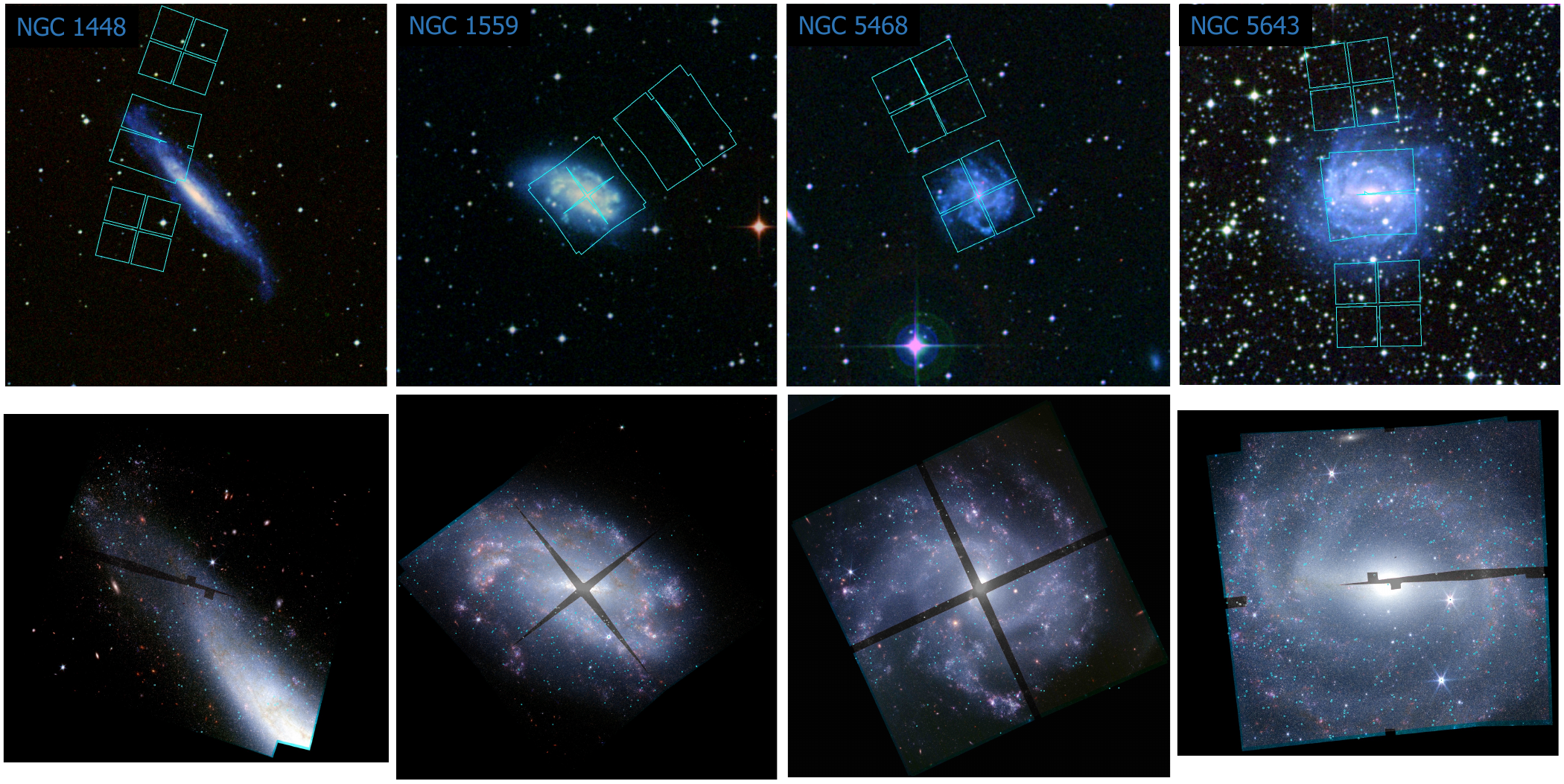}
\end{center}
\caption{\label{fg:images} NIRCam fields superimposed on Digitized Sky Survey color images for four hosts (top) and NIRCam RGB images ({\it F090W/F150W/F277W}) showing positions of Cepheids (cyan circles) (bottom). North is up and east is to the left. }
\end{figure}
\section{Data}

The targets of this program, GO 1685, were selected at those richest in Cepheids and SN Ia from the SH0ES host sample of 37.  The first two to execute, NGC 4258 and NGC 5584, were presented in R23 and the four presented here executed after, in mid 2023. The observational parameters of our {\it JWST} NIRCam imaging campaigns for four SN~Ia hosts (NGC$\,$1448, NGC$\,$1559, NGC$\,$5468 and NGC$\,$5643) are provided in Table~\ref{tb:obs} and shown in Fig.~\ref{fg:images} (see \citetalias{Riess:2023} for the same information for NGC$\,$4258 and NGC$\,$5584).  As in \citetalias{Riess:2023}, the four new hosts were imaged with three filters ({\it F090W}, {\it F150W}, {\it F277W}, centered at 0.9, 1.5 and 2.8$\mu$m, respectively; see Fig.~2 in \citetalias{Riess:2023} for filter curves) in two epochs separated by 15-16 days, with one short wavelength module of the NIRCam instrument covering the host and the other placed on a far halo field as shown in Fig.~\ref{fg:images}.  The epoch separation was selected by the JWST schedulers subject to the requirements of a 15-30 day spacing and an orientation difference within 5 degrees.  For the closer two hosts, NGC$\,$1448 and 5643, we swapped the A and B modules on the host on consecutive visits and thus doubled the spatial coverage of the far halo field on opposite sides.

\begin{figure}[t] 
\begin{center}
\includegraphics[width=0.99\textwidth]{ 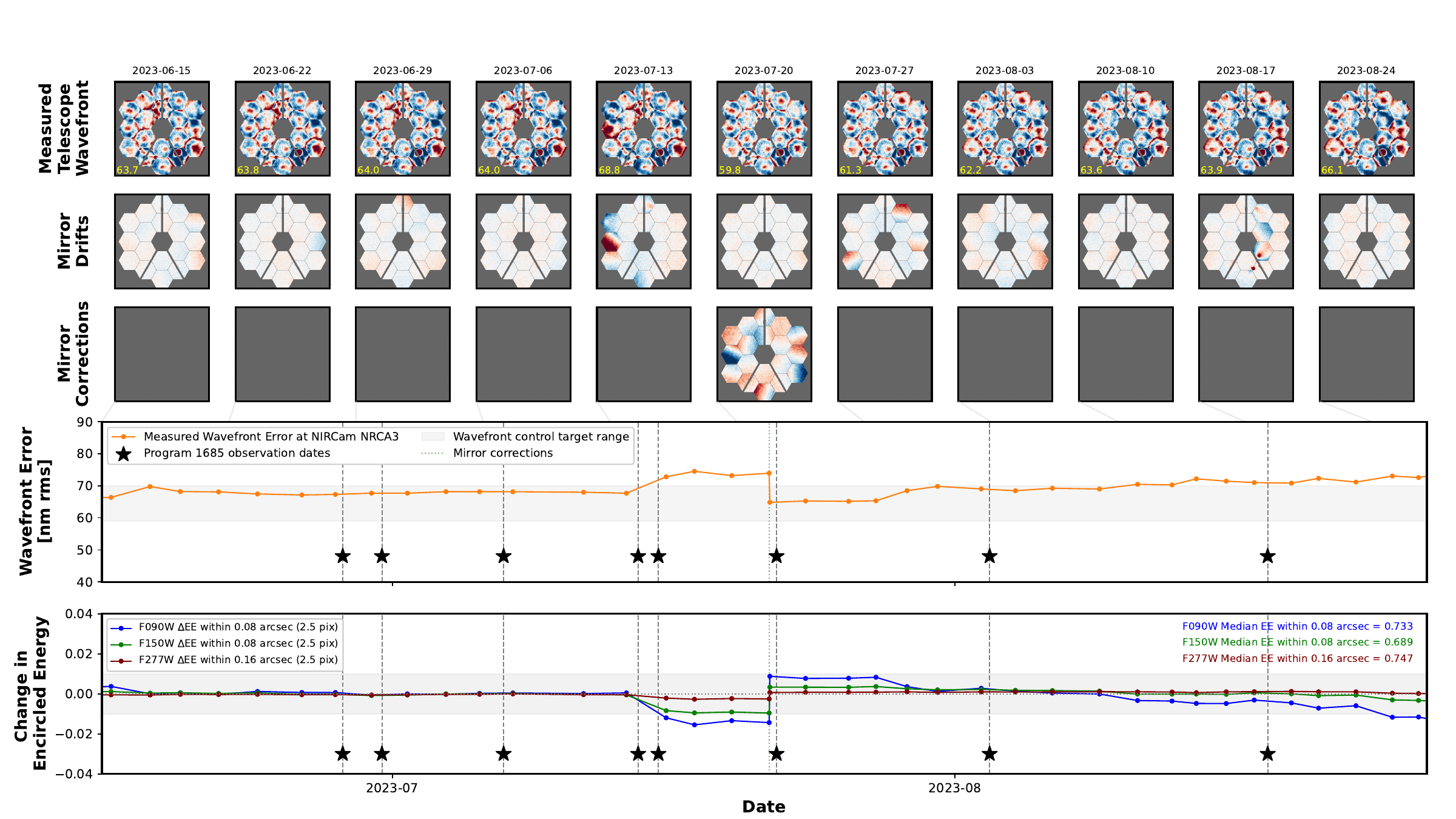}
\end{center}
\caption{\label{fg:wavefront} {\it JWST} wavefront sensing data during the interval of these observations, June-August 2023.  The wavefront sensing, drifts, rectifications and wavefront errors are shown in the top three panels.  The bottom panel shows the change to encircled energy in the core of the PSF in the three filters used for this program.  The dates of our observations are indicated as vertical dashed lines and a star symbol.}
\end{figure}

We analyzed the wavefront history of {\it JWST} over the time span of the observations from the telescope monitoring data, June-August 2023, as shown in Fig.~\ref{fg:wavefront} to determine the photometric stability of our data set.  There was a moderately large tilt of mirror segment C5 detected on July 16 (not present on July 14) that was corrected July 22 and we had no observations between these two dates. The wavefront modeling indicates no large changes at the time of the observations and that variations in encircled energy at small radii (similar to PSF photometry) would be $<$0.005 mag (less at longer wavelengths).  Based on this we judged the epochs to be photometrically consistent and proceeded with our measurements.  We note the impact to photometry due to changes in the shape of the PSF between epochs are further negated by the determinations of aperture corrections from stars within each frame.  

We used the STScI NIRCam reduction pipeline, version 1.12.0, to calibrate the data frames.  There has been only one significant update to the reference files since those used in \citetalias{Riess:2023}, 1125.pmap and 1126.pmap, which included zero point updates for each SCA and flat field improvements.  The present work uses this update (and we have remeasured NGC$\,$4258 and NGC$\,$5584 from \citetalias{Riess:2023}). As a result of these updates, the {\it mean} photometry of sources became brighter by $\sim$ 0.03 mag in {\it F090W} and $\sim$ 0.01 mag in {\it F150W}, where the latter band was used for our baseline distance measurements. Much of this change cancels when comparing Cepheids {\it between} NGC$\,$4258 and NGC$\,$5584.  Updates after 1126.pmap through December 2023 do not affect NIRCam photometry.  

\begin{figure}[b] 
\begin{center}
\includegraphics[width=0.99\textwidth]{ 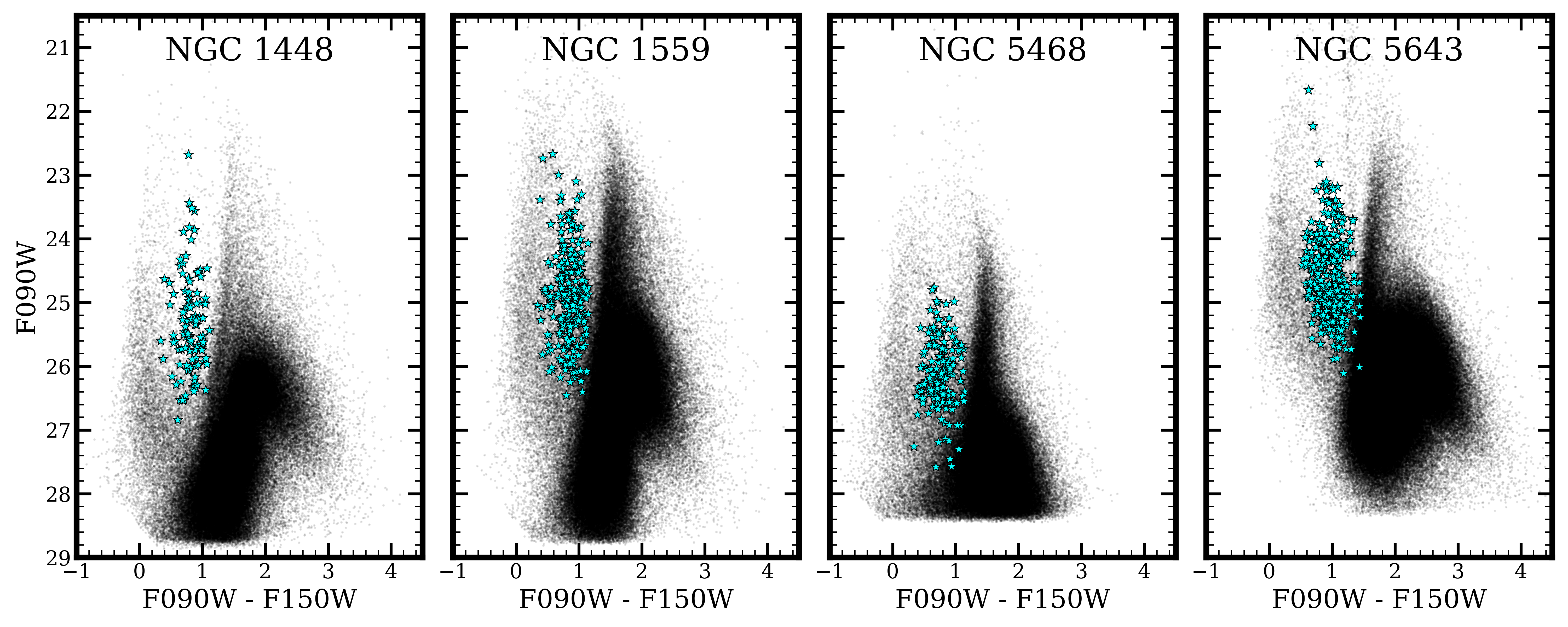}
\end{center}
\caption{\label{fg:cmds} Color magnitude diagrams for the modules covering Cepheids.  Each star is plotted using a small black point, while Cepheids are indicated by cyan symbols.  A color cut in this space is included for NGC$\,$1559, 0.3$<${\it F090W}$-${\it F150W}$<$1.15, corresponding to a broad range around the instability strip (0.5$<${\it V}$-${\it I}$<$1.7). For NGC$\,$5643, which has large foreground extinction ($E(V-I)=0.21$) the color range is shifted redward. The magnitudes and colors of non-Cepheids are based on the first visit. If a Cepheid is observed in only the first or in both visits, its location on the CMD is based on the first visit. Otherwise, it is based on the second visit.}
\end{figure}

We performed photometry on the images using the {\it DOLPHOT} package \citep{Dolphin:2000,DOLPHOT2016} and its NIRCam module \citep{Weisz:2023, Weiszinprep}, with the same procedures described in \citetalias{Riess:2023} with the same cuts on the crowding, sharpness, object type, S/N, and error flags as reported by DOLPHOT\footnote{A revised version of the DOLPHOT package became available after this work was completed, which includes a new ``-etctime" option to revise the exposure times in the image headers.  As a test we compared the Cepheid photometry of NGC$\,$5584 with the old and new version and measured a median difference of 0.001 mag in {\it F150W} and 0.002 mag in {\it F090W} and an 0.3\% mean change in the noise.  We judged these differences too small to merit a change in versions for this work.} (see also \citealt{Warfield2023}).  The only relevant changes since \citetalias{Riess:2023} are the version of the instrumental reference files provided by STScI stated above.  The Cepheids were identified by astrometrically matching the {\it JWST} photometry catalogs to the Cepheid lists identified by the SH0ES team from multi-epoch, optical {\it HST} imaging \cite{Yuan:2022a, Riess:2022}.  The matching tolerance was set to 0.7 NIRCam SW pixels in distance ($0\farcs02$) but matches were generally within $0\farcs01$ of the expected position.  This is the same result as seen in \citetalias{Riess:2023}.  

Among the 36 images we obtained for this program (3 filters in 2 epochs for 6 hosts), only one suffered an issue.   This was for the first epoch in {\it F090W} for NGC$\,$5468 where the number of samples ``up the ramp'' during accumulation (four) resulted in incomplete cosmic-ray removal in the individual \texttt{cal} files (dithers).  This produced some scattered artifacts in that single image.  Although these artifacts are rare enough to have little impact on the PSF-fitting region of the Cepheids, they impacted the measurement of the aperture correction for this image due their presence in source-subtracted sky annuli (which extend to greater radii and thus cover more area than the PSF-fitting region).  To solve this problem, we determined the appropriate aperture correction for this image and in each chip by comparing the non-variable stars between this epoch and epoch 2 with {\it F090W} (which had more samples and no issues).  The result from comparing the non-variable stars between these two epochs produced an aperture correction consistent in value with the set of aperture corrections seen for other images.

\begin{figure}[ht] 
\begin{center}
\includegraphics[width=0.95\textwidth]{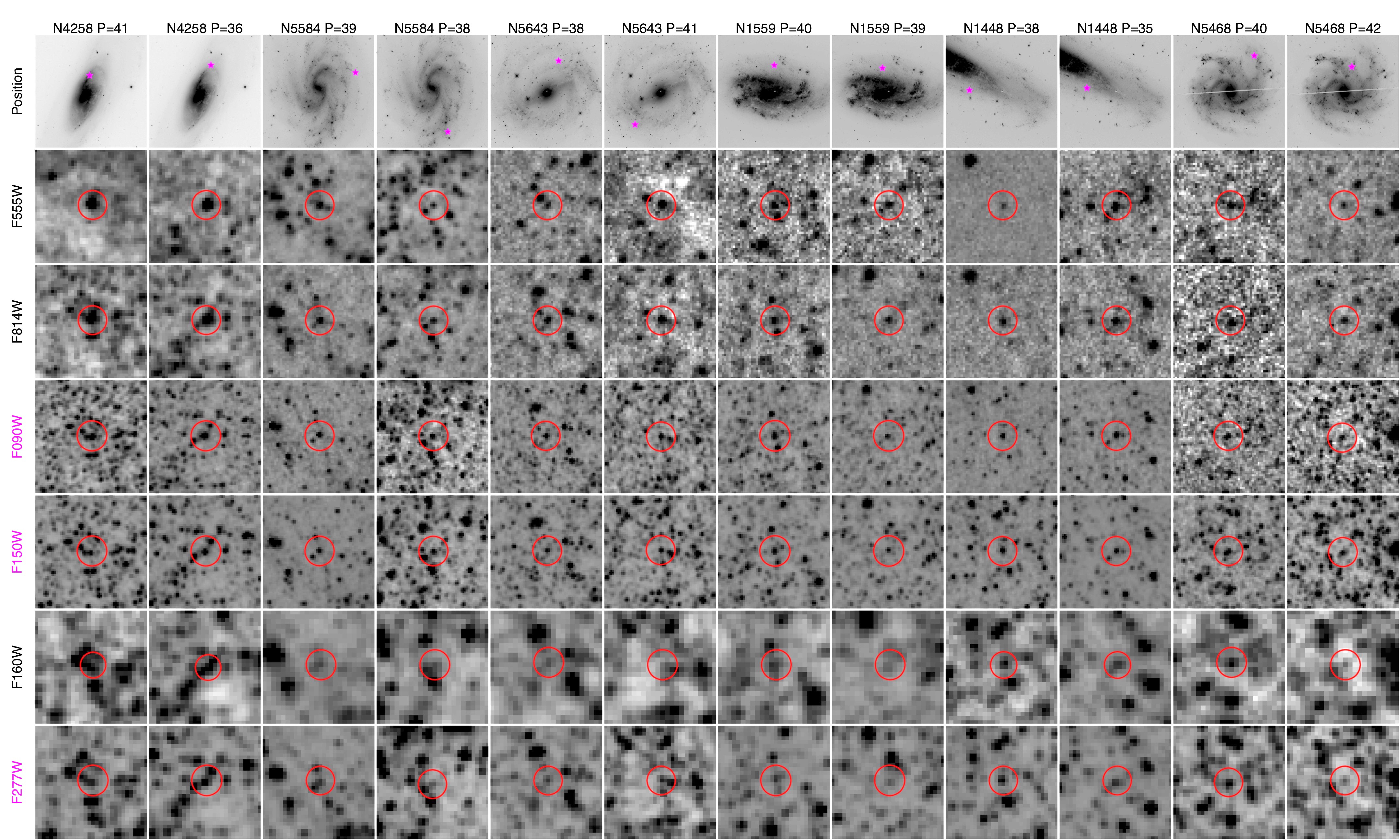}
\end{center}
\caption{\label{fg:stamps} Example {\it HST} and {\it JWST} image stamps around Cepheids with $P \sim$ 40 days in all hosts. The top row shows the location of each Cepheid.  {\it HST} filters are labeled in black while the {\it JWST} filters are in magenta.}
\end{figure}

\startlongtable
\begin{deluxetable*}{lrrrrrrrrrrrr}
\tabletypesize{\scriptsize}
\tablewidth{0pt}
\tablenum{2}
\tablecaption{Photometric Data for Cepheids\label{tb:phot}}
\tablehead{\multicolumn{1}{c}{Host} &  \multicolumn{1}{c}{ID} & \multicolumn{1}{c}{RA} & \multicolumn{1}{c}{Dec} & \multicolumn{1}{c}{log P} & \multicolumn{1}{c}{{\it F090W}} & \multicolumn{1}{c}{$\sigma$}  & \multicolumn{1}{c}{{\it F150W}} & \multicolumn{1}{c}{$\sigma$} & \multicolumn{1}{c}{{\it F277W}} & \multicolumn{1}{c}{$\sigma$} & \multicolumn{1}{c}{{\it V}$-${\it I}$\,^a$} & \multicolumn{1}{c}{$\sigma$}}
\startdata
N5584 &        96196 &  215.58141  &  -0.38763  &  1.2434  &  26.175  &  0.115  &  25.549  &  0.111  &  25.367  &  0.140  &  0.957  &  0.117   \\
N5584 &       114600 &  215.58182  &  -0.39029  &  1.2317  &  26.505  &  0.130  &  25.765  &  0.129  &  25.641  &  0.159  &  0.897  &  0.158   \\
N5584 &       115209 &  215.58295  &  -0.38981  &  1.2015  &  26.598  &  0.138  &  25.940  &  0.173  &  25.783  &  0.198  &  0.817  &  0.157   \\
N5584 &       134727 &  215.58605  &  -0.39115  &  1.2538  &  26.364  &  0.159  &  25.480  &  0.186  &  25.245  &  0.207  &  1.126  &  0.189   \\
\hline
\enddata
\tablecomments{$a$: {\it F555W}$-${\it F814W}.  We note the provided magnitudes from JWST are phase corrected.}
\phantom{}
\vspace{-48pt}
\end{deluxetable*}

As in \citetalias{Riess:2023}, we imposed additional quality requirements on specific photometric parameters to ensure our sample contains reliable Cepheid magnitudes that yield robust distance measurements.  The most important of these is based on the value of $\chi^2$ (the quality of the scene modeling) reported by {\it DOLPHOT}, which we generally require to be better (lower) than 1.4 per degree of freedom; we allow a modest rise in this limit, up to 1.7, as a linear function of log period, since long-period Cepheids are very bright and their lower shot noise will reveal more imperfections in the PSF model.  This cut excluded a median of $\sim$ 7\% of sources among the set of hosts due to subpar scene modeling.  The full Cepheid sample forms a tight locus when comparing log period to $\chi^2$, such that the poor fits are readily apparent as a tail in the $\chi^2$ distribution toward high values. The nature of poor $\chi^2$ objects is that they are either confused even at {\it JWST} resolution or, more likely, include a resolved source such as a cluster or background galaxy, which is not well-modeled with a set of PSFs by DOLPHOT. We also employ the same color cut as \citetalias{Riess:2023}, 0.3$<${\it F090W}$-${\it F150W}$<$1.15 (equivalent to 0.5$<${\it V}$-${\it I}$<$1.7, a broad range for Cepheids) which excludes $\sim$ 5\% of sources.  Nearly all of these are redder than the cut and are either highly reddened or strongly blended with a red star.   Not surprisingly, this red boundary corresponds to the blue edge of the highly populated RGB/AGB branch (see Fig.~\ref{fg:cmds}) so the odds of a direct blend or mis-identification will rise rapidly near this limit with hundreds to thousands of red stars at similar brightness for every Cepheid.  The exception to the above is NGC$\,$5643, the only host with moderate Milky Way foreground extinction ({\it E(V}$-${\it I)}=0.21 mag versus $<$0.05~mag for all others), shifting all apparent colors redward including the RGB/AGB branch and leading us to shift the accepted color range accordingly.  The values of $\chi^2$ for Cepheids in this host are also slightly higher than the other hosts leading us to relax the $\chi^2$ limit by 0.3, resulting in the exclusion of a similar fraction as the other hosts.  

In Fig.~\ref{fg:cmds} we show the position of the Cepheids within the color magnitude diagrams for the field stars in each host.  In Fig.~\ref{fg:stamps} we show cutouts for two representative Cepheids with $P\sim40$ days from each of six host in three {\it HST} and three {\it JWST} filters.  Inspection of the {\it JWST F150W} stamps shows a qualitative change relative to {\it HST F160W} thanks to the higher angular resolution. The background is effectively resolved, with the brightness fluctuations in the {\it HST} images transformed to reveal individual stars and spatially constant backgrounds. 

\begin{figure}[t] 
\includegraphics[width=0.53\textwidth]{ 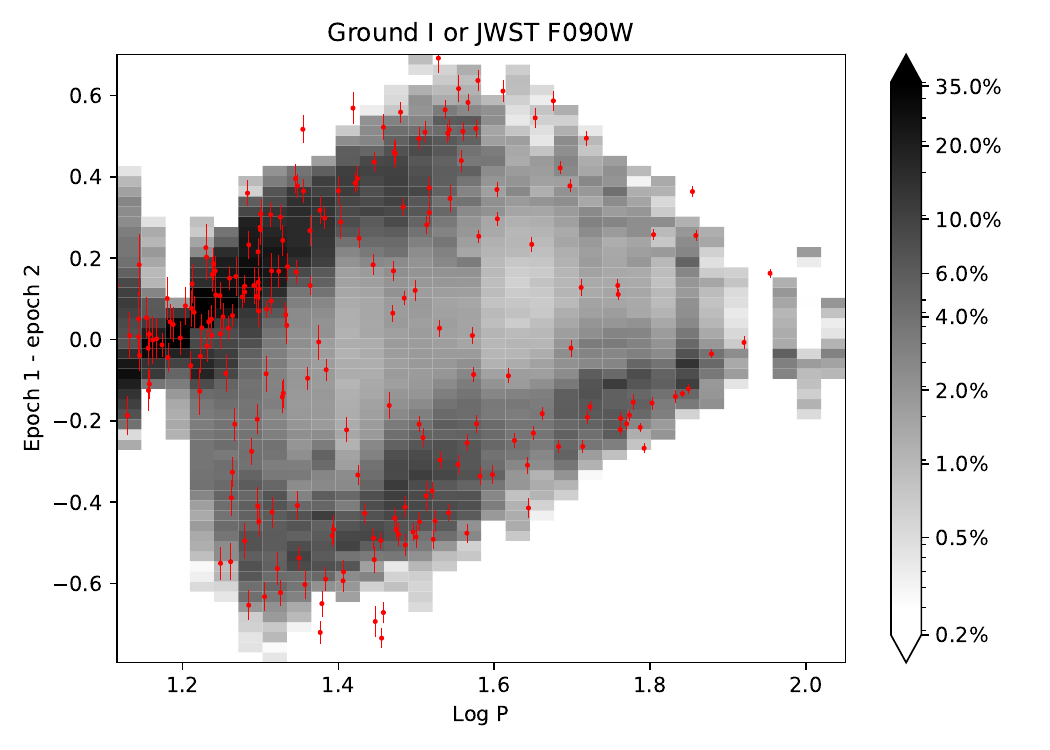}
\includegraphics[width=0.45\textwidth]{ 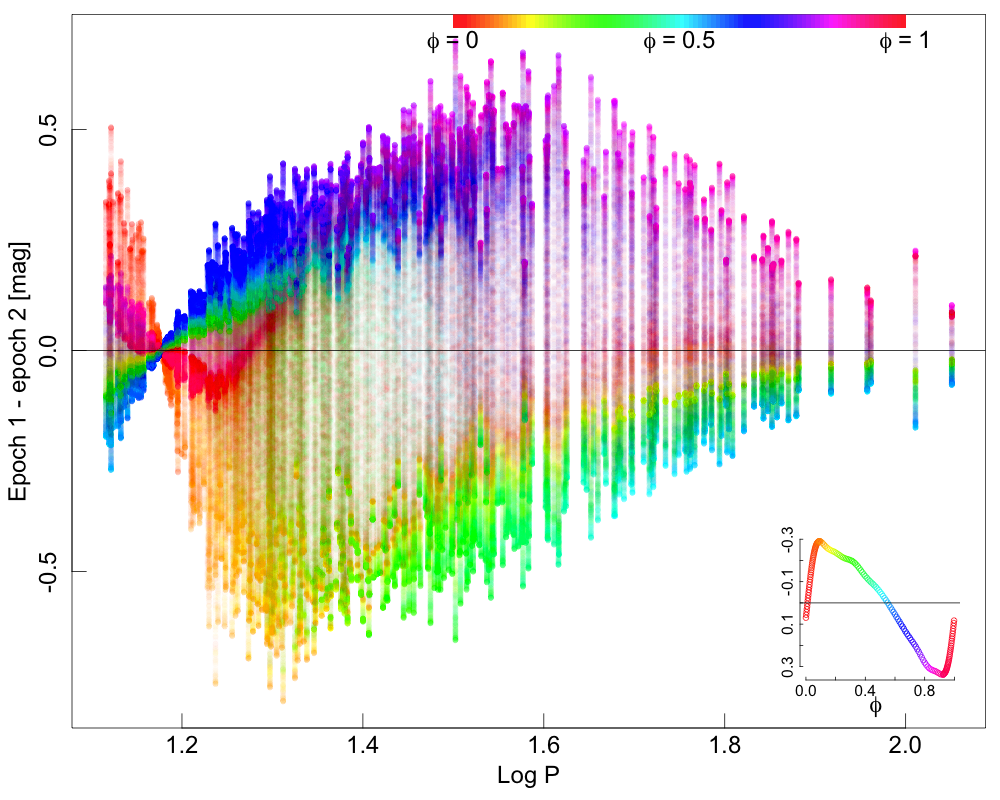}
\caption{\label{fg:phase} Magnitude differences of two epochs for Cepheids in NGC$\,$5643 in {\it F090W} (red points, left) and simulated (gray left and indicating phase, right).  Gray density shows expected frequency of sampling based on random phase and template light curves.  Red points are measurements from {\it JWST} and the errors do not include the crowding error which effectively cancels in the difference.  The asymmetry of Cepheid light curves in {\it F090W} produces structure in this diagram that can be used to constrain the phase. The log of the time interval between epochs produces a negligible difference at a value of $\log P=$1.34.}
\end{figure}

\subsection{Artificial Stars}

As in \citetalias{Riess:2023}, we inject 200 artificial stars that bracket the range of the Cepheid magnitudes, interpolating between these to derive the background bias (i.e., the mean crowding correction) based on the uncrowded magnitudes, estimated from Cepheid periods and an iterative fit to the \PL relation.  These crowding corrections represent the difference, in units of source magnitudes, between the measurement of the source on a uniform background and on the speckled background, determined statistically from the level of nearby sources. The results from these measurements were similar to the two galaxies in \citetalias{Riess:2023} with a mean across all hosts of 0.04, 0.07 and 0.09 mag for {\it F090W}, {\it F150W}, and {\it F277W}, respectively.  These corrections are approximately {\it seven times smaller} than the same quantities for {\it HST} in {\it F160W}.  The random errors derived from the artificial stars ranged from 0.06 mag (NGC$\,$1448) to 0.14 mag (NGC$\,$5468) in {\it F090W} and 0.09 mag (NGC$\,$1448) to 0.16 mag (NGC$\,$5468) in {\it F150W}.  These distributions are slightly asymmetric in {\it F090W} and fairly symmetric in {\it F150W} and {\it F277W}. 

\begin{figure}[t]  
\includegraphics[width=\textwidth]{ 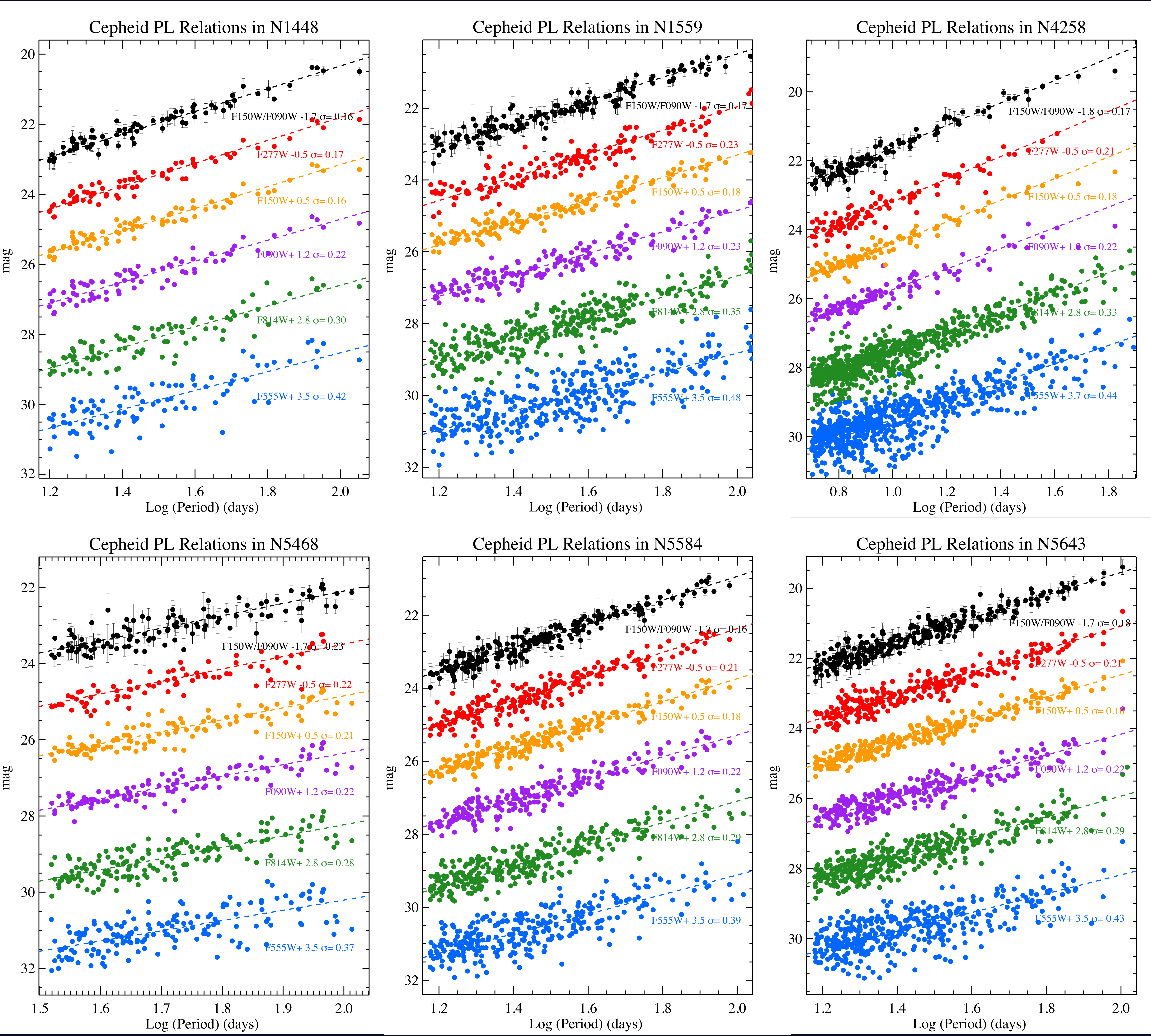}
\caption{\label{fg:pls} \PL relations for four new SN~Ia hosts -- NGC$\,$1448 (top left), NGC$\,$1559 (top center), NGC$\,$5468 (bottom left) and NGC$\,$5643 (bottom right) and for the two galaxies presented in \citetalias{Riess:2023} -- NGC$\,$4258 (bottom right) and NGC$\,$5584 (bottom center). Relations are plotted from bluest (bottom) to least reddened (top). The bottom two relations in each panel are from {\it HST} while the others are from {\it JWST}.  The top relation in each panel, plotted in black, is a dereddened or Wesenheit magnitude, {\it F150W}$-0.72${\it (F090W-F150W)}. Magnitude offsets are applied as indicated for ease of view and the dispersion for each \PL is given.}
\end{figure}
\clearpage

\section{Analysis}

The Cepheids we report on were discovered between 5 and 20 years ago by the SH0ES team using $\sim$12 epochs of optical imaging from {\it HST}, providing a measurement of their periods, amplitudes and photometry in WFC3 {\it F555W}, {\it F814W}, and {\it F160W} \citep{Hoffmann:2016,Riess:2022}.  However, the typical period uncertainty is $\sim$ 2\% \citep{Yuan:2021}, so knowledge of their phases elapses after just a few of years, and certainly by the time of our {\it JWST} observations.  In order to recover their phases we analyzed the change in the Cepheid magnitudes between the two epochs following the methods given in \citetalias{Riess:2023}; we show an example of the relation between the change in magnitude and phase for the Cepheids in one of our hosts, NGC$\,$5643, in Fig.~\ref{fg:phase}.  The phase uncertainty for individual Cepheids is a function of the difference in phase (determined by the Cepheid period) for the two epochs.  We note that the difference photometry between two epochs has less noise than the sum of the two epochs because some of the sources of error, such as the crowding term, are correlated between the two epochs.  The determination of phase (discussed in \citetalias{Riess:2023}) also provides a magnitude uncertainty that takes into account the quality of the phase constraint.

The phase-corrected photometry is provided in Table 2 and includes the combined error terms (from artificial stars, shot noise, empirical determination of the phase, and intrinsic width of the instability strip). We show monochromatic \PL relations for each measured filter in Fig.~\ref{fg:pls}.

\subsection{Reddening-corrected Period-magnitude Relation\label{sec:wes}}

Although reddening in the near-infrared is small compared to the optical, it is not negligible.  Therefore we make use of three dereddened (or Wesenheit) magnitude systems chosen to make the best use of the {\it JWST} data.  These are derived from combinations of 6 filters, three {\it HST} WFC3 bands, {\it F555W}, {\it F814W} and {\it F160W} \citepalias{Riess:2022} and three {\it JWST} NIRCam bands, {\it F090W}, {\it F150W}, and {\it F277W}.  The relations are: 

\begin{itemize}
    \item {\it JWST+HST} NIR, (baseline): $m_H^W=$ {\it F150W}$-0.41$({\it F555W}$-${\it F814W})
    \item {\it JWST} NIR: {\it F150W}$-0.72$({\it F090W}$-${\it F150W})
      \item {\it JWST} MIR: {\it F277W}$-0.30$({\it F090W}$-${\it F150W})
     \end{itemize}

By design these minimize extinction and temperature width effects in the instability strip, aiding in distance measurements by using a color term, $R$, which we derive from the \cite{Fitzpatrick:1999} reddening law ($R_V=3.3$, see \citealt{Brout:2023} for further explanation). The SH0ES team previously used an {\it HST}-only formulation of the baseline relation, $m_H^W=${\it F160W}$-0.39(${\it F555W}$-${\it F814W}$)$; the {\it JWST}+{\it HST} NIR system is very similar but crucially substitutes {\it F150W} for {\it F160W}. This change allows for the most direct comparison with past {\it HST} measurements, reducing NIR confusion noise in past {\it HST} data while leaving the wavelengths measured constant. The color {\it F555W}$-${\it F814W} is well-measured due to the high resolution of {\it HST} WFC3-UVIS and strong contrast between Cepheids and red giants in the optical as seen in Fig.~\ref{fg:stamps}.  The low value of $R$ demonstrates that {\it F150W} is subject to only modest extinction.   We also analyze two other filter combinations which are independent of the {\it HST} measurements of color.  The {\it JWST} MIR system further reduces the impact of extinction by referencing the Cepheids to 2.8$\mu$m.

The mean slope of the $m_H^W$ \PL relation has been well measured with {\it HST} to lie in the range of $-3.26$ to $-3.30$~mag/dex \citep{Riess:2016,Riess:2019a,Riess:2022}.  The mean of these 6 hosts from the {\it HST} measurements is -3.26 $\pm 0.05$. In principle we expect a slightly shallower slope by $\sim 0.01$~mag/dex when substituting {\it HST} {\it F160W} ($\lambda_{\rm eff}=1.53\mu$m) for {\it JWST} {\it F150W} ($\lambda_{\rm eff}=1.50\mu$m) due to the larger color term.  The strongest constraints on the slope come from the LMC due to its large measured period range and low dispersion \citep{Riess:2019a}.  For our baseline we use $-3.25$, near the mean of the {\it JWST} sample which is -3.21 $\pm 0.03$ and the {\it HST} constraint and propagate an uncertainty of 0.05~mag/dex in the slope to the summary results.  We also include variants of the fits which set the slope to $-3.20$ and $-3.30$.

 \subsection{Baseline Results}  

\begin{figure}[t] 
\begin{center}
\includegraphics[width=\textwidth]{ 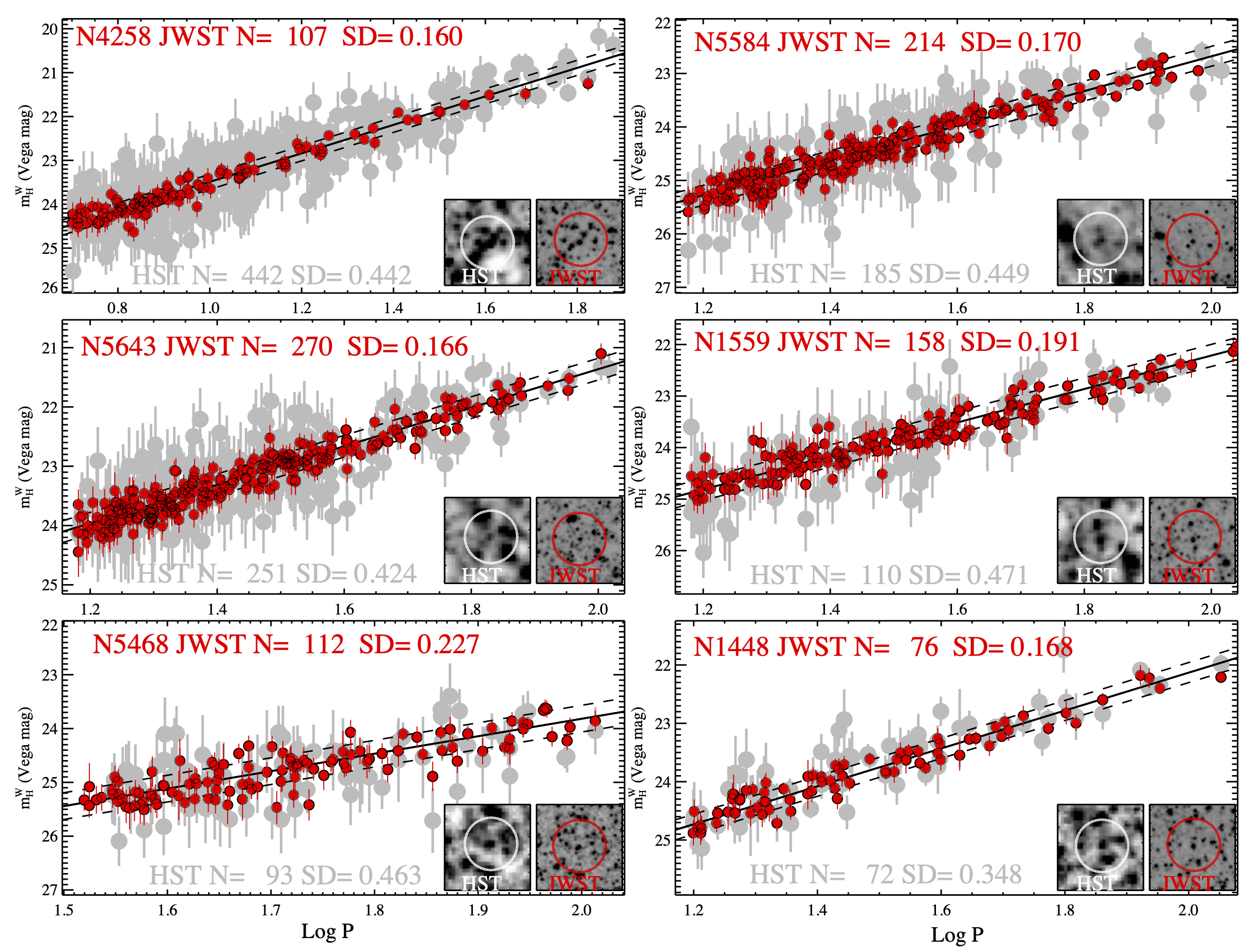}
\end{center}
\caption{\label{fg:comphstjwst} Comparison between the standard (SH0ES: \citetalias{Riess:2022}) dereddened magnitude $m^H_W$ period-magnitude relation used to measure distances to SN~Ia hosts.  The red points use {\it JWST} {\it F150W} ($\lambda_{\rm eff}=1.50\mu$m) and the gray points are from {\it HST} {\it F160W} ($\lambda_{\rm eff}=1.53\mu$m), including a small transformation {\it F150W}$-${\it F160W}$=0.033+0.036[(V-I)-1.0]$ to account for the differences in these passbands.}
\end{figure}

In Table 3 we provide the fits to the Cepheid \PL relations in each host for the baseline system with the NIR measurements from {\it JWST} versus {\it HST} (SH0ES).  
We fit a common formulation,
\begin{equation}
zp=m_H^W-b_W\, (\log\,P-1)-Z_W {\rm [O/H]}, \label{eq:cephmagalt}
\end{equation}
where the zeropoint or intercept, $zp=\mu_0+M^W_{H,1}$, where $\mu_0$ is the distance modulus and $M^W_{H,1}$ is the absolute magnitude of a Cepheid with $\log\,P=1$.  The term $Z_W$ is the Cepheid metallicity dependence in this system \citep[$-0.21$~mag/dex,][]{Breuval:2022, Riess:2022} and we provide the product of this times the difference from solar metallicity, {\rm [O/H]}, in Table 3 for each host.  Because the Cepheids have near-solar metallicity, with measured $ \rm [O/H] \sim 0 $, these metallicity corrections are very small; their typical value is $ \sim 0.01\hbox{--}0.02 $ mag.  We include them for consistency when comparing with previous results from {\it HST}.  We also include a variant where we set the metallicity term to zero and with double the nominal term.

We determine the intercepts within the {\it JWST+HST} NIR magnitude-system \PL relations from the weighted mean after applying an iterative 3$\sigma$ clip (set by Chauvenet's criterion) which removes $\sim$ 3\% of sources ($\sim$30 Cepheids out of $\sim$1000, a comparable fraction to past studies such as \citetalias{Riess:2022}). 
We note that the empirical rejection is applied to the full Cepheid sample discovered in the optical, so this is lieu of that imposed by the {\it HST} NIR data in \citetalias{Riess:2022}.
Because our uncertainties are well-determined and non-uniform we apply rejection as an individual Cepheid contribution to the \PL, $\chi^2 > 3^2$.
We also provide results with no rejection.  Table~\ref{tb:res} provides summary results for the full sample including the mean \PL dispersion.  The small dispersion for the {\it JWST} relations makes the small number of outliers quite evident, as on average they are $\sim$5$\sigma$ off the \PLs.  

\begin{figure}[t] 
\begin{center}
\includegraphics[width=\textwidth]{ 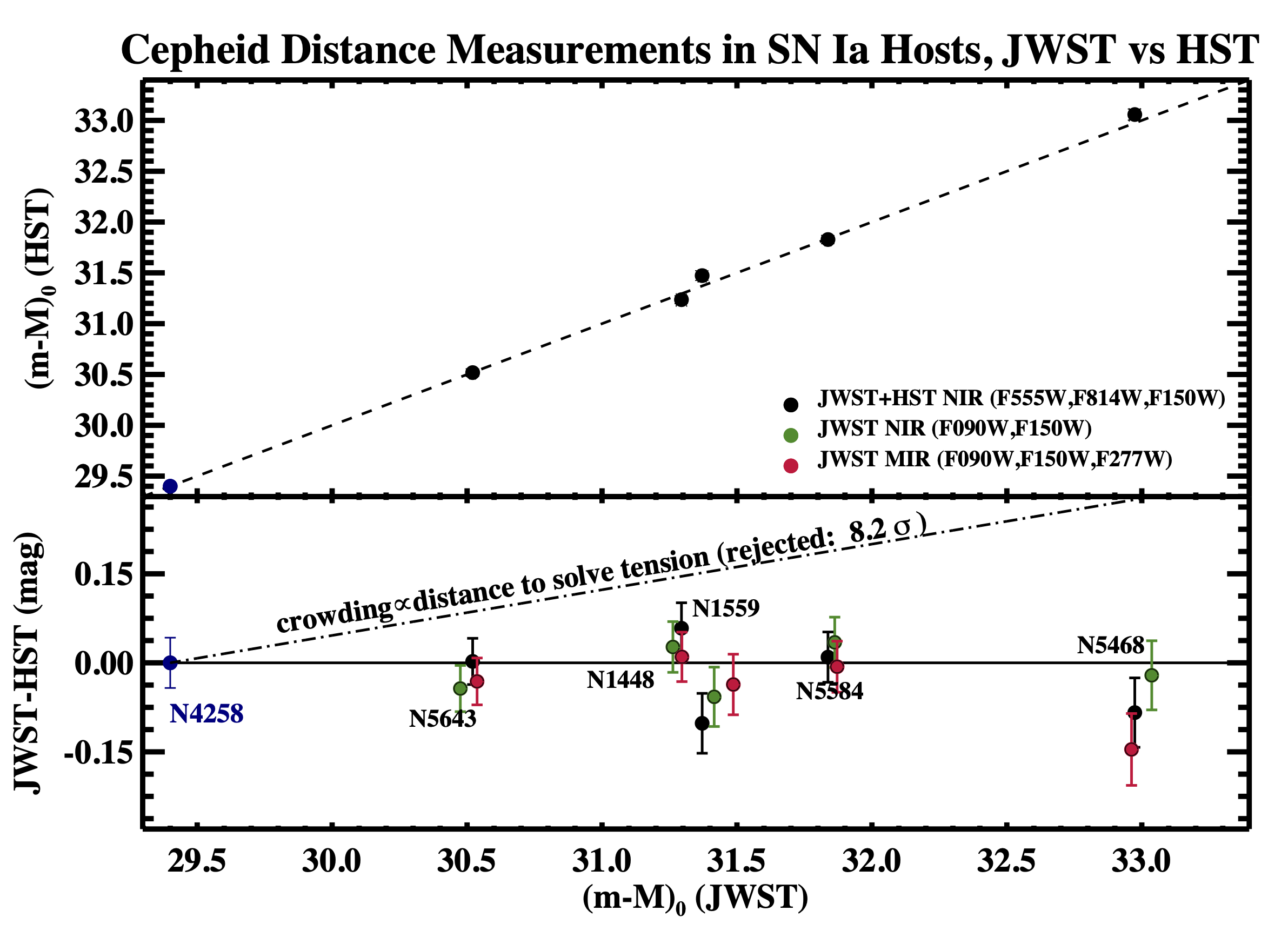}
\end{center}
\caption{\label{fg:summary} Comparison of distances to the five SN~Ia hosts measured with {\it HST} and {\it JWST} anchored by the same geometric distance reference, NGC$\,$4258.  The lower plot shows the differences in the measurements from the two telescopes.  Black shows the comparison for the baseline system used to measure H$_0$, $W^H_{V,I}$ and is the only system plotted on the top panel.  Green and red shows comparisons with two JWST-only magnitude systems.  The bottom plot shows a hypothetical, linear model of unrecognized crowding tuned to match the Hubble Tension, 5log(73/67.5)=0.17 mag at the mean distance of the SH0ES sample, $\mu=31.7$, a trend of 0.07 mag per magnitude of distance modulus beyond NGC$\,$4258.  This model is ruled out at 8.2$\sigma$.}
\end{figure}

The fits are remarkably tighter than those measured with {\it HST} as seen in Fig.~\ref{fg:comphstjwst}, an expectable (but still impressive!) direct consequence of the improved telescope resolution (see \citetalias{Riess:2023}, Fig.~1 for sources of noise in \PL relations).  We see a consistent reduction in dispersion by a factor of 2.5 to a mean of $\leq$0.18 mag, with the three closest at 0.16-0.17 mag.  Only NGC 5468 has a dispersion $>$ 0.2 mag, a consequence of its 1.5 mag greater distance and relatively shorter exposure time.  We determine distances to the SN hosts by using the geometric distance determination of NGC$\,$4258 \citep{Reid:2019} $\mu_{0,N4258}=29.397$~mag and the intercept difference between the SN hosts and NGC$\,$4258, i.e.,

\begin{equation}
\mu_{0,SN}=zp^{SN}-zp^{N4258}+\mu_{0,4258}.
\end{equation}
fit from the \PL relations.  The measured distances from the baseline systems are given in Table 3 with differences between {\it HST} and {\it JWST}, also plotted in Fig.~\ref{fg:summary}.  The baseline mean difference, {\it JWST}$-${\it HST}, seen in Table~\ref{tb:res}, is $-0.011\pm0.032$~mag\footnote{If this difference is interpreted as a mean error in {\it HST} measurements, the sense would be of slightly {\it over-correcting} {\it HST} Cepheid photometry for crowding and underestimating H$_0$. However, the difference is not significant.}. This uncertainty receives nearly equal contributions from three terms of size $\sim$0.02 mag: the intercepts of NGC$\,$4258 measured with {\it HST} and with {\it JWST}, and the mean of the SN host PLs.  The latter term comes mostly from the {\it HST} intercept uncertainties, which are 3 times the size of the {\it JWST} means.  The similarity of the NGC$\,$4258 \PL error terms results from the combinations of smaller dispersion for {\it JWST} (0.16 vs.~0.44~mag) balanced by the smaller {\it JWST} sample ($N=107$ vs.~$N=442$; more fields observed with {\it HST}) which may be remedied in the future with more {\it JWST} pointings for NGC$\,$4258.

\begin{figure}[t] 
\begin{center}
\includegraphics[width=0.65\textwidth]{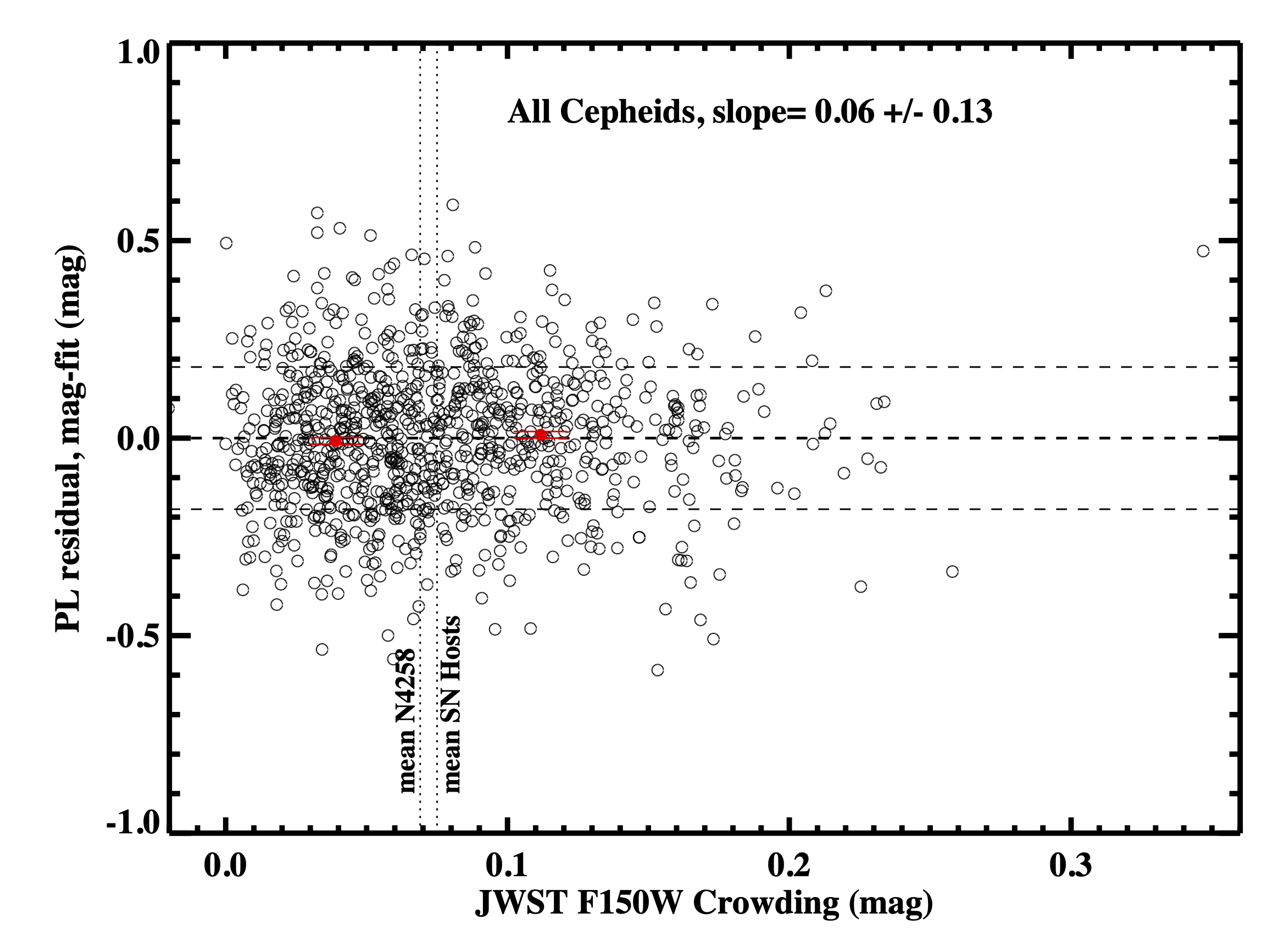}
\end{center}
\caption{\label{fg:crowding} Relation between the $F150W$ \PL fit residuals and crowding, as measured in {\it JWST F150W} frames as the difference between artificial star input and extraction.  We find no dependence between these quantities. Red points and errors shown the mean results if dividing the sample in half.  The mean level of crowding in NGC 4258 and the SN hosts are indicated by vertical lines. The scatter is evenly split as coming from phase uncertainty, crowding (at {\it JWST} resolution) with smaller contributions from the width of the instability strip.}
\end{figure}

Regressing all {\it JWST} crowding corrections versus the residuals from their best fit \PL relations (see Fig.~\ref{fg:crowding}), we find no interdependence between these quantities. The best-fit value of 0.06$\pm$0.13 mag (residual) /mag (crowding) implies that at the mean crowding of the {\it JWST} sample ($\sim$ 0.07 mag) the mean \PL bias is 0.004$\pm$0.008~mag.  This determination provides a useful estimate of the level of error that would apply when comparing this Cepheid photometry to that of Cepheids in the closest hosts, such as the Milky Way and the LMC, where crowding is absent.

We can attempt to gain additional leverage in the {\it JWST} vs.~{\it HST} comparison by factoring in the distances of the hosts.  For this purpose it is useful to devise a hypothetical model in which unrecognized crowding in Cepheid photometry measured with {\it HST} resolution grows linearly with distance (modulus) beyond the calibration from NGC$\,$4258 to cause the Hubble Tension, i.e., a size of $5\log(73/67.5)=0.17$~mag at the mean distance modulus of the SH0ES Cepheid sample ($\mu_0=31.7$~mag), equivalent to $\sim$0.07 mag of bias per magnitude of distance modulus.  This model is shown in Fig.~\ref{fg:summary}.  We can reject this model at 8.2$\sigma$ with the highest leverage from the farthest host, NGC$\,$5468 ($\mu_0=33.0$), which shows no evidence of such an effect.  
{\it Indeed, the evidence we see against crowding as the source of the Tension is now greater than the evidence of the Tension itself.}  As is, we see no evidence of a growing difference between {\it HST} and {\it JWST} with distance as would be required for such a model.

\begin{deluxetable}{lrr|rrrrrr|rrrrrr|rr}[b]
\tablenum{3}
\tablecaption{ Baseline {\it HST}  and {\it JWST} Cepheid \PL Fits  \label{tb:fits}}
\tablewidth{0pt}
\tablehead{\multicolumn{3}{c}{ } & \multicolumn{6}{c}{JWST+HST NIR} & \multicolumn{6}{c}{HST NIR (SH0ES)} & \multicolumn{2}{c}{JWST-HST} \\[-0pt]
\multicolumn{1}{c}{Host} & \multicolumn{1}{c}{metal} & \multicolumn{1}{c}{slope} & \multicolumn{1}{c}{{\scriptsize N}} & \multicolumn{1}{c}{{\scriptsize{zp}}} & \multicolumn{1}{c}{{\scriptsize $\sigma$}} & \multicolumn{1}{c}{{\scriptsize SD}}  &\multicolumn{1}{c}{{\scriptsize $\mu$}}  & \multicolumn{1}{c}{{\scriptsize $\sigma^c$}} & \multicolumn{1}{c}{{\scriptsize{N}}} & \multicolumn{1}{c}{{\scriptsize ZP}} & \multicolumn{1}{c}{{\scriptsize $\sigma$}} & \multicolumn{1}{c}{{\scriptsize{SD}}} & \multicolumn{1}{c}{{\scriptsize $\mu$}}  & \multicolumn{1}{c}{{\scriptsize $\sigma^c$}}  & \multicolumn{1}{c}{{\scriptsize $\Delta$}}  & \multicolumn{1}{c}{{\scriptsize $\sigma$}}}
\startdata
\tableline
n4258 &  -0.018 & -3.25  &  107  &  26.739  &  0.017  &  0.160  & -- & -- &  442  &  26.739  &  0.017  &  0.442  & -- & -- & -- & -- \\
n5584 &  -0.022 & -3.25  &  214  &  29.178  &  0.011  &  0.170  &  31.838  &  0.020  &  185  &  29.168  &  0.032  &  0.449  &  31.828  &  0.037  &  0.010  &  0.042  \\
n5643 &  0.029 & -3.25  &  270  &  27.861  &  0.011  &  0.166  &  30.520  &  0.020  &  251  &  27.859  &  0.028  &  0.424  &  30.518  &  0.033  &  0.002  &  0.039  \\
n1559 &  0.003 & -3.25  &  158  &  28.712  &  0.015  &  0.191  &  31.371  &  0.023  &  110  &  28.813  &  0.041  &  0.471  &  31.473  &  0.045  &  -0.102  &  0.050  \\
n5468 &  -0.020 & -3.25  &  112  &  30.316  &  0.019  &  0.227  &  32.975  &  0.026  &   93  &  30.398  &  0.049  &  0.463  &  33.058  &  0.052  &  -0.083  &  0.058  \\
n1448 &  -0.022 & -3.25  &   76  &  28.630  &  0.017  &  0.168  &  31.289  &  0.024  &   72  &  28.577  &  0.029  &  0.348  &  31.236  &  0.034  &  0.053  &  0.042  \\
\enddata
\tablecomments{\citetalias{Riess:2022} Table 6 $\mu_{N5584}=31.772\pm0.052$ based on three anchors and $P>18$ days; here we use only one anchor, NGC$\,$4258, and $P > 15$ days to allow a direct comparison. To allow a direct comparison with JWST$+$HST NIR, we applied a transformation of $F150W-F160W=0.033+0.036(F555W-F814W-1)$ which adds 0.03-0.04 for most hosts, and corrected here for CRNL by the subtraction of 0.035~mag for all hosts, the two corrections canceling to $< 0.01$~mag with no impact on distance. $^a$: error does not include geometric distance uncertainty for NGC$\,$4258 of $\pm 0.032$.}
\vspace{-36pt}
\end{deluxetable}

\subsection{Baseline Variants}

In Table~\ref{tb:res} we provide the previously discussed variants to the measurement process as well as additional variants which exclude the phase correction, use only Cepheids with {\it HST} {\it F160W} observations (the WFC3-IR field of view is smaller than the optical WFC3-UVIS field where the Cepheids were found and which reduces the sample size), and split the sample into the more and less crowded halves (where the DOLPHOT crowding parameter is above or below the sample median).  This latter test is meant to mimic the sub-selection of Cepheids by those that appear visually least crowded with {\it JWST} resolution, as proposed in \cite{Freedman:2023} for host NGC 7250 (see also Appendix).  This produces a significant difference in sample mean dispersion, namely $\sim$0.17 mag for the less crowded half and $\sim$0.20 mag for the more crowded half, but only a 0.015$\pm0.053$ mag difference in H$_0$ between them.  Although the difference is small, this selection has the potential to bias the remaining sample because it is enacted on the appearance of the source, which may include unresolved blending rather than the statistical properties of the scene.  For example, a coincidental superposition of a source with the Cepheid would {\it appear} to be uncrowded and yet would suffer a bias whose size would be seen with artificial star measurements.  Another likely consequence of imposing a selection is the skewing of distributions of measurements relative to the expected errors as suggested in the values of $\chi^2$.  We caution that any sub-selection of the Cepheid sample needs to be simulated and included in artificial star selection to avoid bias in either measurements or errors \citepalias[see further discussion in][]{Riess:2023}.  

We also determined the results for a single epoch (the first); this provides useful information on the dispersion of the P-L relation when no phase information is available.  The single-epoch solution yields SD=0.224 mag versus SD=0.178 for two epochs and phase corrected (plus an additional 100 Cepheids with only a single epoch), a 25\% reduction in dispersion or equivalent weight to a 70\% increase in the sample size (58\% from the scatter and 12\% more objects).  With perfect measurements, a single, random phase at this wavelength will produce $\sim$ 0.15 mag scatter, two random phases reduces this to $\sim$0.11 mag and two, phase-corrected mags to $\sim$ 0.075 mag.  We note that most of these variants are not statistically independent of each other, as they use the same Cepheid samples and measurements. As all variants and tests yield results consistent with the baseline, we consider the baseline results robust. 

We also provide the summary results from the other two filter combinations based purely on {\it JWST} measurements, {\it JWST} MIR and {\it JWST} NIR as shown in Figure \ref{fg:summary}.  These yield very similar results as the baseline. 

\begin{deluxetable*}{llrcrrrcc}[t]
\tablenum{4}
\tabletypesize{\normalem}
\tablecaption{Results \label{tb:res}}
\tablewidth{0pt}
\tablehead{\multicolumn{1}{c}{Sample} & \multicolumn{1}{c}{Comment} & \multicolumn{1}{c}{\PL} & \multicolumn{1}{c}{$\sigma$-} & \multicolumn{1}{c}{Cepheids} & \multicolumn{1}{c}{SD} & \multicolumn{1}{c}{$\chi^2_{\nu}$} & \multicolumn{1}{c}{\it JWST} & \multicolumn{1}{c}{$\sigma$}\\
& & slope & clip & & & & $-${\it HST} & }
\startdata
JWST+HST NIR  & Baseline &  -3.25  &     3  &   938  &  0.178  &  1.0  &  -0.011  &  0.032  \\
\tableline
JWST+HST NIR  & $\sigma$-clip &  -3.25  &  no  &   966  &  0.218  &  1.7  &  0.017  &  0.031  \\
JWST+HST NIR  & no min P or  $\sigma$-clip  &  -3.25  &  no  &  1005  &  0.222  &  1.7  &  0.015  &  0.031  \\
JWST+HST NIR  & $P > 15$ days &  -3.25  &     3  &   870  &  0.181  &  1.1$^a$  &  -0.020  &  0.043  \\
JWST+HST NIR  & shallower slope &  -3.20  &     3  &   938  &  0.179  &  1.1  &  -0.010  &  0.032  \\
JWST+HST NIR  & steeper slope &  -3.30  &     3  &   937  &  0.179  &  1.1  &  -0.008  &  0.032  \\
JWST+HST NIR  & no metallicity cor. &  -3.25  &     3  &   937  &  0.178  &  1.0  &  -0.005  &  0.032  \\
JWST+HST NIR  & double metal cor. &  -3.25  &     3  &   939  &  0.179  &  1.0  &  -0.016  &  0.032  \\
JWST+HST NIR  & and in SH0ES F160W  &  -3.25  &     3  &   611  &  0.172  &  1.1$^a$  &  0.014  &  0.035  \\
JWST+HST NIR  & no phase correction &  -3.25  &     3  &   941  &  0.191  &  1.0$^a$  &  -0.007  &  0.033  \\
JWST+HST NIR  & lower half {\it JWST} crowding &  -3.25  &     3  &   487  &  0.165  &  1.1$^a$  &  -0.023  &  0.035  \\
JWST+HST NIR  & higher half {\it JWST} crowding &  -3.25  &     3  &   451  &  0.201  &  0.9$^a$  &  -0.006  &  0.040  \\
JWST+HST NIR  & first epoch only &  -3.25  &     3  &   821  &  0.224  &  1.0$^a$  &  -0.007  &  0.036  \\
\tableline 
JWST NIR & F090W,F150W &  -3.25  &     3  &   929  &  0.177  &  1.0  &  -0.012  &  0.032  \\
JWST MIR & F090W,F150W,F277W &  -3.25  &     3  &   864  &  0.197  &  1.0  &  -0.030  &  0.033  \\
\enddata
\tablecomments{$a$: These variants effect the errors as well as magnitudes or sample so $\chi^2_{\nu}$ not directly comparable.}
\end{deluxetable*}

\section{Discussion}

The sample of observations of $\sim$1000 Cepheids with {\it JWST} in 5 hosts of 8 SNe~Ia and in NGC$\,$4258 provides {\it very} strong evidence that NIR {\it HST} Cepheid photometry is accurate, albeit noisier than from {\it JWST}.  While this may not be surprising since past observations used artificial stars to correct for crowding, the search for an explanation of the Hubble Tension merits a broad array of investigations including independent checks on Cepheid photometry.  

Although the sample of Cepheids remeasured with {\it JWST} and presented here represents nearly a third of the full SH0ES sample, we refrain from providing a value of H$_0$ determined exclusively from the {\it JWST} data because the sample size of calibrated SN~Ia and independent geometric anchors is substantially inferior to what is available in \citetalias{Riess:2022}.  The true value of the {\it JWST} data provided here is to provide a high-fidelity test of the {\it HST} Cepheid measurements.

Additional {\it JWST} observations of some Cepheids in the SH0ES host sample are available from other Cycle 1 programs.  These observations are more difficult to compare directly with {\it HST} observations because of the respective program design; however, they appear broadly consistent with the conclusions found here and in \citetalias{Riess:2023}.    \cite{Yuan2022JWST} presented a comparison of Cepheids in NGC$\,$1365 from program GO-2107 (PI Lee) with NIRCam {\it F200W}. They obtained a similar result as the current program, albeit with lower significance, partly because the observations were not ideal for this purpose due to the very different wavelength than {\it HST} {\it F160W} and limited depth.  We analyzed the first set of observations by program GO-1995 (PI Freedman), the only non-proprietary set at the time of writing, of the SH0ES host NGC$\,$7250 observed in $F115W$ and $F444W$ and discussed in \cite{Freedman:2023}.  NGC$\,$7250 is one of the smallest hosts in the sample in terms of physical size and has the lowest mass and the {\it HST}-based analysis by \cite{Riess:2022} includes only 21 Cepheids, about 8 times less than the average host in this work.  Consequently, the comparison between {\it HST} and {\it JWST} for NGC$\,$7250 is limited in precision by the $\pm 0.13$ mag error from the {\it HST} intercept.   The comparison is further limited by the lack of more than one epoch to determine phase corrections for {\it JWST} and the large difference between {\it JWST} and {\it HST} filters ($\lambda_{\rm eff}=$1.15 vs.~1.53$\mu$m) necessitating a larger photometric transformation of $\sim$0.4$-$0.5 mag, which at present can only be obtained from model SEDs.  Nevertheless, a comparison of the {\it JWST} and transformed {\it HST} observations of NGC$\,$7250, presented in the Appendix, are consistent at the $\sim$1.1$\sigma$ level, with {\it JWST} brighter than {\it HST} (the direction of decreasing the distance to the host and increasing H$_0$).  
\subsection{Outlook}

The Cepheid measurements from {\it HST} have passed a very strong test of their accuracy provided by the resolution of {\it JWST}.  At this point,  a solution to the Hubble Tension is most likely to exist elsewhere, because the evidence against a bias in {\it HST} Cepheid photometry is greater than the evidence of the Tension itself.  

We anticipate related improvements afforded by the remarkable capabilities of {\it JWST}.  Specifically, we expect additional calibrations of SNe~Ia from enhancements to the primary distance indicators of TRGB, JAGB, and Miras as well as Surface Brightness Fluctuations.  We suggest the most effective manner of comparing these distance indicators is by using them to measure the distance {\it to the same set of hosts} rather than by comparing values of H$_0$ which necessarily involves additional rungs and potentially unrelated differences. Tying all of these together by observing large samples in common can lead to the calibration of $\sim$ 100 SNe Ia and a $<$1\% local measurement of H$_0$, a landmark in our quest to understand the expansion of the Universe.

\begin{acknowledgments}
    We are indebted to all of those who spent years and even decades bringing {\it JWST} to fruition. This research made use of the NASA Astrophysics Data System.  We thank Martha Boyer, Yukei Murakami and Siyang Li for helpful conversations related to this work.  We thank our PC Alison Vick.  We thank an anonymous referee for improving the draft.

Some of the data presented in this paper were obtained from the Mikulski Archive for Space Telescopes (MAST) at the Space Telescope Science Institute. The specific observations analyzed can be accessed via \dataset[DOI].

We acknowledge support from {\it JWST} GO-1685.  RIA is funded by the SNSF via an Eccellenza Professorial Fellowship PCEFP2\_194638 and acknowledges support from the European Research Council (ERC) under the European Union's Horizon 2020 research and innovation programme (Grant Agreement No. 947660).
This research was supported by the Munich Institute for Astro-, Particle and BioPhysics (MIAPbP) which is funded by the Deutsche Forschungsgemeinschaft (DFG, German Research Foundation) under Germany´s Excellence Strategy – EXC-2094 – 390783311.

\end{acknowledgments}
\clearpage
\appendix

\section{Cepheids in NGC$\,$7250}

\begin{figure}
\begin{center}
\includegraphics[width=0.85\textwidth]{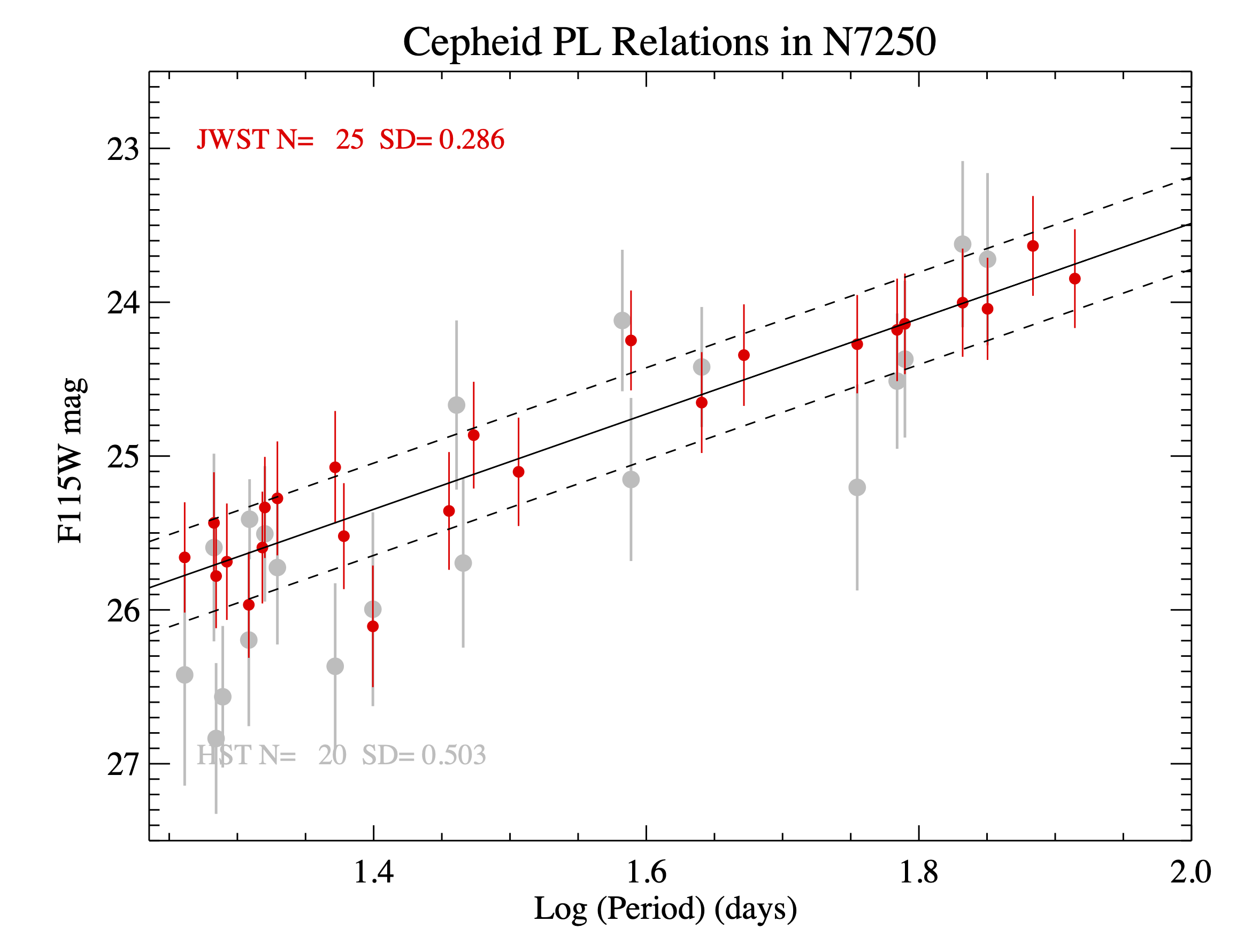}
\end{center}
\caption{\label{fg:7250} Comparison of Cepheid photometry in NGC$\,$7250 in $F115W$.  Gray shows photometry from {\it HST F160W} from \citet{Hoffmann:2016} and \citetalias{Riess:2022}, transformed to {\it F115W} using a synthetic relation, {\it F160W}$=${\it F115W}$+$ 0.41$\pm$0.05$+$0.49({\it F555W}$-${\it F814W}$-1)$.  Points in red are from our measurements of images from {\it JWST} Program 1995 (PI Freedman).  The \PL relations are broadly consistent ({\it JWST} brighter than {\it HST} by 1.1$\sigma$). The {\it JWST} dispersion is a factor of 2 smaller than {\it HST}.}
\end{figure}

We retrieved the {\it JWST} NIRCam observations of NGC$\,$7250 from program 1995 (PI Freedman) from the MAST archive, calibration version 1130.pmap.  We found the Cepheid observations in {\it F444W} (4.4$\mu$m) to be of poor resolution due to the broad PSF and poor contrast between the Cepheids and red giants so we did not analyze them. Cepheids from the sample from \cite{Hoffmann:2016} and \cite{Riess:2022} were well-detected in {\it F115W} and we measured their photometry with DOLPHOT following the same steps discussed above.  Unfortunately a second useful color from {\it JWST} was not available to apply a similar color quality cut as employed here and  only a single epoch was obtained so we could not measure phase corrections.  Also, the {\it JWST} filter {\it F115W} ($\lambda_{\rm eff}=1.15\mu$m) is quite different than the {\it HST} filter  {\it F160W} ($\lambda_{\rm eff}=1.53\mu$m) with no overlap,  so unlike our comparisons with {\it JWST} {\it F150W} presented above, quantifying the difference requires a rather large, SED model-dependent (and hence uncertain) color transformation between the filters, equivalent to inferring the $J-H$ colors of a Cepheid from these models.  We used the Padova isochrone SED models to provide a synthetic relation between colors, {\it F160W}$=${\it F115W}$+0.41 \pm 0.05+0.49$({\it F555W}$-${\it F814W}$-1)$ for Cepheid like SEDs, i.e., typical Cepheids with {\it F555W}$-${\it F814W}$\sim$1 must be offset by $\sim$0.4$-$0.5$\pm$0.05~mag to compare between the telescopes.  
We identified 27 Cepheids in $F115W$ from the optical sample of 40 (with minimum completeness Log P $>1.25$) from \cite{Hoffmann:2016} using the same procedure as in \S 3.  We used a a 2.5$\sigma$ clip (appropriate for this small sample size of $N \leq 40$ by Chauvenet's criterion) for both the {\it JWST} and {\it HST} samples as shown in Fig.~\ref{fg:7250}.  The clip rejected two of the twenty-seven in the JWST set and one of the twenty-one Cepheids in NGC 7250 provided in \cite{Riess:2022}, ID=65093 with $P=82.1$ days.   The 25 that remained for JWST have a dispersion of 0.29 mag.  Given the small Cepheid sample size, larger \PL dispersion, and large filter difference, we can only conclude the \PL relations are broadly consistent ({\it JWST} brighter than {\it HST} by 1.1$\sigma$) and that the {\it JWST} dispersion is a factor of $\sim$2 smaller than {\it HST}.  A more conservative period completeness limit (see \cite{Hoffmann:2016}) of Log P $>1.4$ yields an ever closer match with the remaining {\it JWST} sample fainter by 0.3$\sigma$ but leaves a very small comparison sample of 11 and 16 Cepheids for HST and JWST, respectively. The mean crowding for this irregular, star-bursting Sdm type host is substantially higher due of its compact nature, $\sim$2 times that of the large spiral hosts studied here and comprising most of the \citetalias{Riess:2022} sample.  We also considered a ``least crowded'' sample selection, following \cite{Freedman:2023} who sought to reduce scatter by visual selection of the least crowded Cepheids according to the {\it JWST} images.  As we did in \S 3 for the data from our program, a less crowded sample was selected as those with a DOLPHOT ``crowd'' parameter which was below the full sample median, 0.135.  The reduced sample of 13 Cepheids have a mean crowding of 0.07 mag (comparable to the mean of the six hosts studied here without a crowding cut, estimated for the same 1.15 $\mu$m wavelength to be $\sim$0.05 mag) have a reduced scatter of 0.18 mag and the same intercept as the full sample within $\sim$0.02 mag.  This sample is very small so that minor differences in its composition may produce large, stochastic variations.  However, we are skeptical that there is any real gain to be made by sub-selecting less crowded Cepheids because their lower scatter is already reflected in their smaller uncertainties (measured with artificial stars) and their greater weight.  In Table~\ref{tb:res}, this same selection also yielded a similar result, lowering the dispersion while yielding a difference in mean intercept of 0.01 mag, and given the also larger and independent samples, 40 times larger, the significance of this test is substantially stronger.

\bibliographystyle{apj} %
\bibliography{bibdesk}
\end{document}